# Atomic-level description of thermal fluctuations in inorganic lead halide perovskites

Oliviero Cannelli[1], Julia Wiktor[2], Nicola Colonna[3,4], Ludmila Leroy[1,5], Michele Puppin[1], Camila Bacellar[6], Ilia Sadykov[6], Franziska Krieg[7,8], Grigory Smolentsev[6], Maksym V. Kovalenko[7,8], Alfredo Pasquarello[9], Majed Chergui[1*], Giulia F. Mancini[10*]

[1] Laboratory of Ultrafast Spectroscopy (LSU) and Lausanne Centre for Ultrafast Science (LACUS), École Polytechnique Fédérale de Lausanne, CH-1015 Lausanne, Switzerland.
[2] Department of Physics, Chalmers University of Technology, SE-412 96 Gothenburg, Sweden.
[3] Laboratory for Neutron Scattering and Imaging, Paul Scherrer Institute, CH-5232 Villigen-PSI, Switzerland.
[4] National Centre for Computational Design and Discovery of Novel Materials (MARVEL), École Polytechnique Fédérale de Lausanne, CH-1015 Lausanne, Switzerland.
[5] LabCri, Departamento de Física, Universidade Federal de Minas Gerais, 31270-901 Belo Horizonte, Brazil.
[6] Paul Scherrer Institute (PSI), CH-5232 Villigen, Switzerland.
[7] Institute of Inorganic Chemistry, Department of Chemistry and Applied Biosciences, ETH Zürich, CH-8093 Zürich, Switzerland.
[8] Laboratory for Thin Films and Photovoltaics, Empa-Swiss Federal Laboratories for Materials Science and Technology, CH-8600 Dübendorf, Switzerland.
[9] Chaire de Simulation à l'Echelle Atomique (CSEA), École Polytechnique Fédérale de Lausanne (EPFL), CH-1015 Lausanne, Switzerland.
[10] Department of Physics, University of Pavia, I-27100 Pavia, Italy.
[*]Email: majed.chergui@epfl.ch; giuliafulvia.mancini@unipv.it

**Abstract**

The potential of lead-halide perovskites for optoelectronic and thermoelectric applications is currently hindered by their limited long-term stability under functional activation. While the role of lattice flexibility in the thermal response of perovskites has become increasingly evident, the description of thermally-induced distortions is still unclear.

In this work, we provide a unified picture of thermal activation in $CsPbBr_3$ across length scales, showing that lattice symmetry does not increase at high temperatures - contrarily to previous claims. We combine temperature-dependent X-ray diffraction (XRD), Br K-edge X-ray absorption near-edge structure (XANES), *ab initio* Molecular Dynamics (MD) simulations, and calculations of the XANES spectra by first-principles, accounting for both thermal fluctuations and core hole final state effects. We find that the octahedral tilting of the Pb-Br inorganic framework statistically adopts multiple local configurations over time - in the short-range. In turn, the stochastic nature of the local thermal fluctuations uplifts the longer-range periodic octahedral tilting characterizing the low temperature structure, with the statistical mean of the local configurations resulting in a cubic-like time-averaged lattice. These observations can be rationalized in terms of displacive thermal phase transitions through the soft mode model, in which the phonon anharmonicity of the flexible inorganic framework causes the excess free energy surface to change as a function of temperature.

Our work demonstrates - for the first time - that the effect of thermal dynamics on the XANES spectra can be effectively described for largely anharmonic systems, provided *ab initio* MD simulations are performed to determine the dynamically fluctuating structures, and core hole final state effects are included in order to retrieve an accurate XANES line shape. Moreover, it shows that the soft mode model, previously invoked to describe displacive thermal phase transitions in oxide perovskites, carries a more general validity.

**Introduction**

The implementation of lead-halide perovskites as active medium in photovoltaic and optoelectronic devices is currently hindered by their relatively poor long-term stability[1,2]. Indeed, despite their promising performances[3,4], these materials suffer from degradation under operative conditions, caused by external factors such as oxygen, moisture and light[5–7]. Composition and structural instabilities were also observed, due to ion migration [8], phase segregation [9] and thermal heat effects[10]. Even though effective strategies to mitigate some of these issues are already under development, as in the case of degradation due to moisture[11,12], solutions to other forms of instabilities still require additional investigation. Specifically, thermal and optical stimuli, which represent two forms of functional activation respectively exploited for thermoelectronics[13–15] and optoelectronics[16,17], are also possible sources of degradation.

The perovskite structural instabilities are closely related to the peculiar flexibility of the Pb-X (X=Cl⁻, Br⁻, I⁻) lead-halide framework, characterized by pronounced lability and anharmonicity[18–23]. The deformable lattice results in small free energy formation of anion vacancies and low activation energy for vacancy-mediated transport, implying intrinsic high anion mobility[24] and causing reversible phase separation when multihalide perovskites are exposed to continuous illumination[9].

The study of a photoexcited system requires to disentangle optically-induced electronic effects from possible thermal ones. In our recent work on $CsPbBr_3$ nanocrystals (NCs)[25], we demonstrated that thermal effects are not significant in the photoinduced response, which is characterized by polaronic lattice distortions that we specifically quantified with atomic level precision. Here, we focus on the characterization of the temperature-induced (T-induced) response of the $CsPbBr_3$ NCs in a temperature range relevant for optoelectronic applications.

The structure of $CsPbBr_3$ consists of a Pb-Br inorganic sublattice in which the $Pb^{2+}$ ions are surrounded by 6 $Br^-$ anions, making a framework of corner-sharing octahedra. The cuboctahedral voids left in the crystal are filled by the Cs$^+$ ions, creating a complementary sublattice where the cations undergo a free rattling motion[19]. In single crystals, the phase diagram of this system was determined by neutron and X-ray diffraction (XRD) studies, which identifies phases of increasing symmetry upon temperature rise: orthorhombic $Pnma$ (T < 88 °C), tetragonal $P\frac{4}{m}bm$ (88 °C < T < 130 °C) and cubic $Pm\bar{3}m$ (T > 130 °C) [26,27]. In recent years, this simple description in terms of transformations of ideal lattices across the phase diagram was questioned, even though the presence of structural disorder at high temperatures was already suggested in the seminal work of Møller on inorganic lead-halide perovskites single crystals[28]. Thermal distortions of the cubic unit cells were shown in both methylammonium (MA$^+$) and caesium lead bromide single crystals through first-principles molecular dynamics (MD) and low-frequency Raman scattering[22]. Temperature-induced rotational disorder of lead-halide octahedra was also observed in a high energy resolution inelastic X-ray scattering (HERIX) and pair distribution function (PDF) study on MAPbI$_3$ single crystals at 350 K, ascribing the effect to the phonon anharmonicity of the Pb-I cage[20].

In nanostructured perovskites, for which ligand-capping strategies are often employed to enhance the systems' stability and to influence their thermal and optoelectronic properties[12,29,30], the scenario is even more complex. Polycrystalline NCs are particularly difficult to characterize with conventional X-ray powder diffraction, due to their high surface-to-bulk ratio and the diffuse scattering of the organic ligands[23,31]. In an XRD investigation of CsPbBr$_3$ perovskite NCs, two phase transitions were reported in the temperature ranges T=50-59 °C and T=108-117 °C and ascribed to orthorhombic-tetragonal and tetragonal-cubic phase transitions, respectively[21]. However, approaches beyond the standard Rietveld refinement suggested that structural defectiveness in inorganic lead-halide NCs can be related to the presence of orthorhombic twinned subdomains, present both at room and higher temperatures[23].

Even though a large variety of experimental and computational methods have been employed to investigate the flexibility of the lead-halide structure, a unified picture simultaneously describing T-induced changes in the short-range and in the long-range is still missing. Additionally, the reported static and dynamic local disorder contrast with the assignment of highly ordered phases in the long-range, especially at high temperatures, raising questions about the phase diagrams proposed in early works[26,27,32]. Some studies ascribed the perovskites high temperature phase to an apparent cubic structure resulting from the statistical average of disordered local structures[20,22,23], but the disorder itself was never fully reconciled with the presence of well-defined phase transition temperatures. Conversely, Monte Carlo simulations at finite-temperatures assigned the qualitative features of the phase transitions of CsPbBr$_3$ to pure thermal lattice activation[33], however the authors did not benchmark their theoretical results against experimental observables.

Here, we present a correlative characterization of the short-range and long-range structures in T-activated CsPbBr$_3$ NCs. We perform T-dependent measurements of powder XRD and Br K-edge X-ray absorption near edge structure (XANES) spectroscopy on CsPbBr$_3$ NCs in the 25-120 °C range, i.e. across the phase diagram of the nanostructured material[21], and compare them to *ab initio* MD calculations as a function of temperature. The XRD and XANES observables were computed as averages over several configurations extracted from the MD trajectory, with the XANES spectra additionally including core hole final state effects. This approach goes beyond standard structural refinement methods, and fully accounts for the statistical fluctuations of the lattice structure, which are particularly pronounced in this system.

Our results show that, in the high temperature phase, the system adopts multiple local configurations characterized by dynamically-evolving structural distortions. We quantify the magnitude of these distortions, which have an extent comparable to those of the low-symmetry orthorhombic phase, and show that the statistical mean of these local configurations results in a cubic time-averaged lattice. Moreover, we demonstrate that the longer-range structure results from the statistical average of locally distorted configurations. Hence, the local thermal fluctuations cause the periodic octahedral tilting in the orthorhombic low-temperature structure to break, reproducing the temperature-dependent changes of the XRD experiment. Our observations are a direct consequence of the intrinsic phonon anharmonicity of the lead-halide sublattice, and are rationalized in terms of displacive thermal phase transitions *via* the soft mode model[34].

These results highlight the difference between thermal and light-induced structural responses in CsPbBr$_3$ perovskites, the former being intrinsically random in nature and the latter selectively driven by electron-phonon coupling[25]. The deeper understanding of the perovskite responses upon different stimuli will open new opportunities for manipulating and stabilizing the lattice structure in realistic applications.

**Methods**

Temperature-dependent (T-dependent) XRD and XANES measurements were performed at the SuperXAS beamline at the Swiss Light Source (SLS) of the Paul Scherrer Institute. The concept of the experiment is depicted in Figure 1. The sample consists of a powder of long-chain zwitterion-capped CsPbBr$_3$ dry perovskite NCs with cuboidal shape (side length 11.9±2.2 nm) and high photoluminescence quantum yield[29]. The sample was located in a thermostated cell holder between two 0.254 mm-thick graphite layers, and the internal temperature of the cell was calibrated and monitored throughout the experiment with a thermocouple.

T-dependent XRD measurements were performed using a monochromatic 12.9 keV X-ray beam in transmission geometry, with a sample-detector distance of 24.1 cm. The transmitted diffraction signal was acquired using a Pilatus 100k 2D detector (94965 pixels, 172x172 µm$^2$ pixel area) and then azimuthally averaged in the 1.3-2.8 Å$^{-1}$ Q-range to obtain the radial averaged intensity *I(Q)*:

$$I(Q) = \frac{1}{2\pi} \int I(Q, \varphi) d\varphi$$

where *I(Q, φ)* represents the scattered intensity at the defined scattering vector *Q* and the azimuthal angle *φ*. The background XRD signal generated by the graphite sheets was isolated in a dedicated measurement, without the perovskite sample. Temperature-dependent XANES measurements were conducted using a 5-element silicon drift detector (SDD) for fluorescence detection at 90° geometry. The spectra were collected at the Br K-edge (13.450-13.569 keV) using a crystal Silicon (111) monochromator. At each energy point, the spectra were normalized by the incident X-ray flux. A flat pre-edge offset was subtracted for each spectrum and the intensity was normalized by the absorption edge integral. XRD patterns and XANES spectra were recorded at the temperatures: 25, 35, 40, 45, 55, 60, 65, 120 °C.

*Ab initio* MD simulations based on density functional theory (DFT) were performed using the CP2K package[35]. The Perdew-Burke-Ernzerhof (PBE) functional[36] was used to describe the exchange-correlation energy. Three different MD simulations, lasting for 10-16 ps and using a timestep of 5 fs, were carried out in the isobaric (*NpT*) ensemble. In the runs, the initial shape of the cell was kept constant, while the volume of the cell was allowed to fluctuate. One MD calculation was run at 27 °C (300 K) with the orthorhombic geometry as initial condition. At 130 °C (403 K), two simulations were run, one initialized with the cubic and one with the orthorhombic geometry. These structures were chosen to monitor the thermal dynamics of the system in its lowest and highest structural phases. For the high-temperature simulations, two different starting geometries were considered in order to evaluate the impact of the initial conditions on the computation. Simulations were carried out in supercells containing 1080 atoms, which corresponds to the 6x6x6 repetition of the unitary cubic cell. The Brillouin zone was sampled solely at the Γ point. The first 5 ps of the simulations were considered as equilibration and were excluded from the statistics. The mean XRD *I(Q)* profiles were calculated averaging the scattering intensities predicted by VESTA[37] for instantaneous structures separated by 0.75 ps extracted from the MD trajectories. Three additional MD simulations, with the same parameters and starting conditions, were also performed for smaller supercells (320 atoms, corresponding to the 4x4x4 repetition of the unitary cubic cell) to generate structures for the computationally demanding XANES simulations.

XANES spectra were computed performing first-principles DFT calculations using the Quantum Espresso distribution[38,39]. The exchange-correlation effects were described using the PBE functional[36], and the ultrasoft pseudopotentials from the PS-library[40] were employed to model the electron-ion interaction. Br K-edge spectra were simulated with the XSpectra code[41,42] within the excited-electron plus core-hole (XCH) approximation[43]. Calculations were based on 320-atom structures obtained from *ab initio* MD simulations in the CP2K package[35], corresponding to the 4x4x4 repetition of the unitary cubic cell. MD calculations were run at 27 °C (300 K) for an initial orthorhombic geometry, and at 130 °C (403 K) for initial orthorhombic and cubic geometries. For each MD simulation, 5 structures corresponding to 5 different time delays of the MD trajectory were considered. For each time delay, 10 separate XCH calculations were performed with a core hole placed on a randomly chosen Br site of the supercell, for which the Br K-edge spectrum was computed. The final Br K-edge spectra result from the average of the 10 Br sites in each of the 5 MD structures, for a total of 50 spectra per MD simulation. Additional XCH simulations were performed for 160-atoms supercells which were built starting from either the ideal orthorhombic or cubic unit cells, using the atomic coordinated reported in the literature from PDF refinements at 22 °C and 160 °C, respectively[21]. For each non-equivalent Br site, separate XCH calculations were performed, and the average Br K-edge XANES spectrum was computed. Details about the experimental and computational methods are described in the Supporting Information (SI).

**Results**

$CsPbBr_3$ lattice structures for the ideal cubic and orthorhombic unit cells are depicted in Figure 2a,b. Starting from the cubic symmetry, the orthorhombic phase is obtained distorting the $PbBr_6$ octahedra along both the equatorial and axial planes, causing a deviation of the Pb-Br-Pb angle from the 180° value. Model XRD *I(Q)* profiles for orthorhombic and cubic unit cells are shown in Figure 2c, respectively bottom and top, as predicted in the VESTA software[37] at 12.9 keV X-ray incident energy. The profiles differ for the assignment of the main reflection peaks and for the superlattice features which are present in the orthorhombic XRD profile, labelled as *(i), (ii), (iii)*. The latter arise from the periodic recurrence of cooperative octahedral tilting, which doubles the cubic unit cell constant along the axis perpendicular to the tilting direction[44], as represented in Figure 2b.

The T-dependent XRD *I(Q)* profiles obtained experimentally are shown in Figure 3a. The pronounced peak at 1.827 Å$^{-1}$ scattering vector comes from the background signal generated by the graphite sheets enclosing the sample, and it is superimposed to the main reflections (022)-(202) and the superlattice peak *(ii)* of the orthorhombic phase. The XRD *I(Q)* profiles predicted by *ab initio* MD simulations are reported in Figure 3b for three different starting conditions: 27 °C and orthorhombic symmetry (grey); 130 °C and orthorhombic symmetry (yellow); 130 °C and cubic symmetry (red). MD simulations at 27 °C confirm the presence of the three superlattice peaks *(i)*, *(ii)*, *(iii)*, and indicate that, within the lattice thermal motion, the $PbBr_6$ octahedral tilting from which these features originate is overall preserved in the long-range, coherently with the room-temperature description of Figure 2b. In the XRD experiment (Figure 3a), we observe the progressive suppression of the orthorhombic superlattice peaks *(i)* and *(iii)* (grey shaded areas) upon temperature increase (25-120 °C), with the quality of the NCs being preserved in the entire temperature range. Model XRD *I(Q)* profiles, and literature results[21,26,27], generally ascribe the results in Figure 3a to thermal phase transitions, associated to a symmetry increase of the $CsPbBr_3$ unit cell from the orthorhombic up to the cubic lattice. However, the MD simulations at 130 °C show that a strong decrease of the same superlattice peaks intensities occurs equally for the runs initialized with orthorhombic and cubic structures. The statistical displacement of Cs, Pb and Br sites with respect to their average positions, derived by MD simulations (see SI, Figure S8), shows that the $CsPbBr_3$ lattice is characterized by pronounced local distortions and becomes dynamically more active with temperature. Although the two initial symmetries for the MD simulations at 130 °C are different, the thermal dynamics lead to the same result in both cases, *i.e.* a time-averaged lattice resulting from statistical mean of the local configurations, centred on the highest symmetry position (*i.e.* cubic), with statistically-distributed local distortions of magnitude comparable to those of the orthorhombic phase (see Figure S6). The slight differences in the predictions of the two calculations can be ascribed to the different boundary conditions imposed to the supercell for the two starting symmetries. In fact, in both cases the thermal dynamics lift the original symmetry of the structure and dynamically distorts the lattice. Both MD simulations at 130 °C effectively reproduce the high temperature XRD *I(Q)* profile of our experiment: the disappearance of the *(i), (ii), (iii)* superlattice peaks is therefore due not only to an increase in lattice symmetry (*i.e.* orthorhombic to cubic), but – as demonstrated here – to the presence of sufficient thermal fluctuations, which cause the breaking of the long-range periodicity associated with the tilting of the room temperature orthorhombic structure.

In Figure 4a, we show the Br K-edge experimental XANES spectra obtained at 25 °C (blue full line) and 120 °C (red full line). Above the edge, the XANES spectrum is caused by single and multiple scattering events of the photoelectron emitted by the Br atoms against the neighbouring atoms, and it contains information about the angles and the bond distances between the probed site and its nearest-neighbours[45,46]. Figure 4a shows that the XANES spectrum is affected by the temperature increase both at pre-edge energies[47] and above the edge, which represents the ionization limit. The XANES traces were scaled by their total areas and they exhibit a first peak at the Br edge (the so-called white line, at 13.472 keV), related to the Br 1s-4p electronic transition, followed by post-edge modulations peaked at the energies 13.4875 keV and 13.510 keV. The temperature rise from 25 °C to 120 °C induces an intensity decrease of the main features and an intensity increase of the local minima, with an overall broadening of the spectral features. This is best visualised in Figure 4b where the difference between the 120 °C and the 25 °C experimental XANES spectra is shown in purple. The data shows a broad negative feature in the rising-edge region at energies 13.466-13.478 keV, with a global minimum at the edge position of 13.472 keV. A pronounced modulation is also observed up to 50 eV above the edge, with damped positive and negative features respectively peaked in the local minima and maxima of the steady-state spectrum.

Figure 4a also shows the XANES spectra computed with *ab initio* simulation for MD calculations at 27 °C for an orthorhombic starting geometry (dashed grey) and at 130 °C for orthorhombic and cubic initial configurations (respectively dashed yellow and dashed red). For comparison, we report computations for the two pristine orthorhombic and cubic lattice structures (respectively dotted blue and dotted red). In all simulations, core hole final state effects were included, thus the difference between the XANES spectra of the MD and pristine structures lies on the presence or absence of local thermal fluctuations in the lattice. Specifically, the predictions of the MD simulations correspond to the statistical average of multiple XANES spectra of Br sites in different local environments. Instead, the calculations for the pristine structures reflect the XANES spectra of the averaged room temperature (orthorhombic) and high temperature (cubic) configurations, as obtained from a PDF refinement of XRD data, which preserve the translational symmetry of the lattice in the long- and short-range[21].

All simulations reproduce the Br K-edge main features; intensity deviations with respect to the experiment are due to systematic errors of the calculations which, however, cancel out when performing spectral differences between computed XANES spectra (Figure 4b). The two MD-simulated XANES traces at 130 °C have similar line shapes, both showing a blue shift of the rising edge and an intensity reduction of the main peaks at the energies 13.472 keV, 13.486 keV and 13.506 keV with respect to the MD simulation at 27 °C. A corresponding increase of the XANES intensity occurs at the energy of the local minima, in agreement with the experiment. Conversely, the comparison of the XANES spectra for the pristine cubic and orthorhombic structures shows that the former is significantly sharper than the latter and is characterized by one additional feature at 13.498 keV, in net contrast with the experiment.

The XANES differences for the orthorhombic and cubic MD simulations at 130 °C minus the orthorhombic MD simulation at 27 °C are shown in Figure 4b in yellow and red, respectively. In the same Figure 4b, we also present the curves obtained subtracting the orthorhombic MD simulations at 27 °C from different linear combinations of the XANES spectra for the orthorhombic and cubic MD simulations at 130 °C (color-coded from yellow to red upon an increasing relative weight of the cubic MD contribution). All linear combinations yield a negative dip centred at 13.468 keV, followed by a rise around 13.480 keV and intensity modulations at higher energies. The qualitative agreement with the experiment is very satisfactory, especially the post-edge modulations starting from 13.485 keV, despite the intensity mismatch of the feature at 13.477 keV. The spectral difference of the pristine cubic minus orthorhombic XANES spectra consists in two positive peaks at 13.470 keV and 13.483 keV, and a negative band between 13.4725 and 13.480 keV, followed by smaller post-edge modulations at higher energies. The first three features of this spectrum are completely absent in the experimental difference reported in Figure 4b and are traced back to the structural changes occurring when the translational symmetry of the lattice is preserved on the local scale. Overall, XANES is an essential complementary tool to XRD, as it pins T-induced modulations in the line shape to an increase of the local thermal fluctuations of the perovskite lattice, in agreement with the MD predictions.

**Discussion**

The comparison of the experimental and theoretical XANES spectra in Figure 4 rules out the assumption of a local symmetry increase of the $CsPbBr_3$ lattice upon temperature rise, contrary to previous conclusions[21,26,27]. The simulations point to an increment of the sub-lattice disorder caused by thermal fluctuations, leading to an overall broadening of the main spectral features. Specifically, this finding suggests that a symmetry reduction occurs in the system, rather than a symmetry increase.

These results agree with the recent short-range characterization of the perovskites structures available in the literature. Sub-lattice disorder in the metal-halide framework was observed in organic lead- and tin-halide perovskites[48]. The PDF structural refinements obtained from the Fourier transform of X-ray powder diffraction showed significant internal distortions of the $BX_6$ octahedra (with B=$Pb^{2+}$, $Sn^{2+}$ and X=$Cl^-$, $Br^-$) at short interatomic distances. The presence of orthorhombic twin domains in the high temperature phases of inorganic lead-halide perovskites NCs was identified employing Debye scattering equation analysis in a X-ray total scattering study[23]. In single crystals, thermal local fluctuations of the Pb-Br framework were experimentally shown for $CsPbBr_3$ and $MAPbBr_3$ in their highest temperature phase[22]. A zero-frequency Raman peak was observed, which is normally absent in purely harmonic systems, pointing to the presence of strong anharmonicity in the lead-halide framework. Similar conclusions were proposed for $MAPbI_3$ single crystals using the HERIX technique[20] and in both organic and inorganic lead iodide perovskites based on *ab initio* MD simulations[49]. In light of these findings, we argue that the description of the system's phase diagram in terms of symmetry increasing phases upon temperature rise is too simplistic, because it neglects the substantial importance of the lattice dynamics on the structural and electronic properties of $CsPbBr_3$[50,51], such as the unusual optical band gap increase of lead halide perovskites with the temperature due to octahedral tilting[52,53].

Here, we rationalize our results considering the phonon anharmonicity of the inorganic perovskite framework. Phase transitions can be described by the phenomenological Landau-Ginzburg theory, which defines the temperature dependence of the free energy as a power series in an order parameter[54]. For the interpretation of structural phase transitions in oxide perovskites, which are strongly anharmonic systems, an effective microscopic description called soft mode model was proposed from neutron diffraction studies[55]. In this picture, at least one phonon frequency is substantially affected by temperature changes due to anharmonic effects. This "soft" phonon mode represents the order parameter of the phase transition. Following the renormalized phonon theory, the intrinsic dependence of the soft phonon frequency from the temperature can be espressed as[56]:

$$\widetilde{\omega}_k^2 = \omega_k^2 + \alpha \cdot T$$

where $\widetilde{\omega}_k$ and $\omega_k$ are respectively the renormalized and negative harmonic phonon frequencies of the soft mode with wave vector **k**, respectively, and α is a positive factor including the anharmonicity constants of the system. $\widetilde{\omega}_k^2$ determines the sign of the restoring force of the system against a deformation along the soft mode coordinate. As schematically reported in Figure 5a, if $\widetilde{\omega}_k^2 < 0$, the free energy surface has a double-well shape with a negative curvature at the high symmetry position, and the system is stabilized by a structural distortion along the soft mode. This applies to CsPbBr$_3$, which is in the orthorhombic phase at room temperature.

In Figure 5b, the statistical distribution of the Pb-Br-Pb angle predicted by the MD simulation at 27 °C (orthorhombic starting geometry) is shown in grey. The tilting is reported as the difference between 180° and the Pb-Br-Pb angle projected along the XZ plane, such that any angular distortion with respect to the ideal cubic structure implies a deviation from the 0° value. We note that the distribution is characterized by a wide bimodal shape of the Pb-Br-Pb angle peaked at symmetric positions of ±16°. This result is consistent with the room temperature XRD *I(Q)* profiles reported in Figures 2b,3a, characterized by the superlattice peaks arising from the periodic tilting of the PbBr$_6$ octahedra in the system. The fit of the distribution with two identically symmetric Gaussian curves yields a standard deviation of 12° (details are reported in the SI, Figure S7).

Being α positive, a temperature increase changes the renormalized phonon frequency $\widetilde{\omega}_k^2$ first to zero and then to positive values, stabilizing the cubic configuration of the system since a restoring force acts on the nuclei when they displace from the high symmetry positions[34]. Correspondingly, the minimum of the excess free energy curve is displaced as depicted in Figure 5a. This process is thus defined "displacive" phase transition and it is driven by temperature-dependent anharmonic effects. In Figure 5b, our MD simulations at 130 °C (orthorhombic and cubic initial configurations, in orange and red, respectively) show the impact of the excess free energy modification on the statistical distribution of the octahedral tilting. In fact, upon temperature increase from 27 °C to 130 °C the Pb-Br-Pb angle distribution drastically changes, with the high temperature curves characterized by a broad monomodal distribution centred at the high symmetry position (0°) with a standard deviation of 18°. Thus, even at 130 °C does the statistical weight of strongly distorted configurations remain relevant, with extreme absolute values up to 40-60°, which are similar to those of the orthorhombic structure at 27 °C. As such, the temperature increase mainly affects the centre of the distribution, *i.e.* the distortions of the Pb-Br-Pb bond in proximity of the ideal cubic geometry, in agreement with the predictions of a displacive phase transition. Correspondingly, the superlattice peaks in the high temperature XRD *I(Q)* profiles of Figure 3 disappear due to the loss of long-range periodicity associated with the orthorhombic tilting (Figure 2b), caused by thermal fluctuations and anharmonic effects.

The soft mode model provides a microscopic description of the changes occuring in the system across the phase transition, solving the ambiguities about the presence of structural disorder in the high temperature phase of $CsPbBr_3$[20,22,23]. The soft phonon frequency undergoes a continuous change with the temperature due to the phonon anharmonicity, and so it does the shape of the excess free energy curve. Upon temperature increase, a discontinuity in $\widetilde{\omega}_k$ occurs at the phase transition temperature, defined as the temperature at which $\widetilde{\omega}_k^2$ changes in sign, stabilizing the average high symmetry position. At the critical temperature, other physical properties also undergo a discontinuity, such as the linear thermal expansion coefficient[57] or the ultrasonic velocity[32] for the $CsPbBr_3$ system.

We argue, therefore, that the soft mode model carries a more general validity in describing displacive thermal phase transitions in perovskites, whether oxide or lead-halide and, for the latter, either inorganic or organic. Indeed, a recent T-dependent neutron diffraction study reported the presence of a soft mode in $MAPbBr_3$ single crystals[58], with the results also being interpreted in terms of displacive thermal phase transitions driven by the $PbBr_6$ octahedral tilting. Due to the common Pb-Br framework of organic and inorganic lead bromide perovskites, we can ascribe the consecutive phase transitions in $CsPbBr_3$ to T-dependent anharmonic effects of the $PbBr_6$ tilting mode, with the soft mode representing the order parameter of the inorganic perovskite system. The proposed scenario is also consistent with recent observations reported in[51].

Long-range structural techniques probe the average lattice geometry of the system, which in the highest temperature phase is centred in the cubic symmetry positions. However, thermal dynamics determine strong local distortions of the lattice, which are observed using short-range structural characterization methods, such as XANES spectroscopy used here, where the pristine cubic cell is inadequate to reproduce the spectrum of the high temperature phase. Since the XANES signal originates from the statistical average of all local configurations of the probed sites, this flexible system cannot be precisely described with a unique averaged configuration representative for all local geometries, at high temperatures. To our knowledge, the effect of thermal disorder on the XANES spectra was successfully reproduced only for systems in which the harmonic approximation of the phonon modes is appropriate[47,59]. Therefore, here we demonstrate that the impact of thermal dynamics on the XANES spectra can be effectively described also for largely anharmonic systems such as lead-halide perovskites when *ab initio* MD simulations - to determine the dynamically fluctuating structures - are combined with calculations accounting for core hole final state effects to retrieve an accurate XANES line shape. Previous works characterized different sources of disorder at the local scale in the perovskite systems, but mostly relied on *a-posteriori* strategies in which a single time-averaged structure was considered, proposing only qualitative arguments to reconcile their inconsistency with longer range results, which agree with high symmetry structures[21,23,27,48].

**Conclusions**

In this work, we provide the consistent microscopic description of the thermal dynamics of $CsPbBr_3$ combining short- and long-range structural sensitive techniques with *ab initio* MD simulations. Based on the agreement between theory and experiments, we harness our first-principles results to precisely quantify the thermal fluctuations of the system, retrieving unprecedented details on the temperature-dependent structural changes.

The atomic-level picture emerging from the correlative characterization of $CsPbBr_3$ is very different from the light-driven structural changes observed in $CsPbBr_3$ NCs with time-resolved XANES[25]: upon above band gap excitation, large polarons are formed due to the electron-phonon coupling between the photocarriers and the polar inorganic lattice. The crystal distortion involves the activation of one specific longitudinal-optical phonon mode, implying well defined nuclear displacements of the Pb-Br sublattice, which can be retrieved with atomic scale precision by the analysis of the transient spectra. Instead, the high temperature configuration of the system is dynamically distorted and cannot be reduced, at the local scale, to an average ordered structure, thereby questioning the classical picture of a low-to-high symmetry phase transition.

These results clarify the underlying mechanisms of the lattice response under functional activation and offers strategies to control the perovskite nuclear degrees of freedom with different external stimuli. Understanding the thermal processes acting at the atomic level represents the first step towards a rational design of perovskite-based devices with improved stability.

**Data Availability**

Processed data showed in this manuscript are available in the Supplementary Information. Raw data are available in the repository: 10.5281/zenodo.4928984

**Associated Content**

Supporting information: (1) Samples and characterization. (2) T-dependent XRD and XANES data analysis. (3) MD computational methods. (4) XANES computational methods. (5) Time evolution of the Pb-Br-Pb angle distribution in the MD simulations. (6) Thermal displacements of the Cs, Pb, Br sites in the MD simulations.

**Author Information**

Notes: The authors declare no competing financial interests.

**Acknowledgments**

This work was supported by the European Union's Horizon 2020 research and innovation program, through the grant agreement no. 851154 (ULTRAIMAGE) and the no. 695197 (DYNAMOX). J.W. acknowledges funding from the Swedish Research Council (2019-03993) and the Chalmers Gender Initiative for Excellence (Genie). The computations were performed on resources provided by the Swedish National Infrastructure for Computing (SNIC) at NSC, C3SE, and PDC. N. C. acknowledges the support of by the SwissNSF NCCR-MARVEL. M.K. acknowledges funding by the European Union's Horizon 2020 program, through a FET Open research and innovation action under the grant agreement no. 899141 (PoLLoC). We thank Balázs Őrley for the graphical rendering of Figure 1.

**Figures**

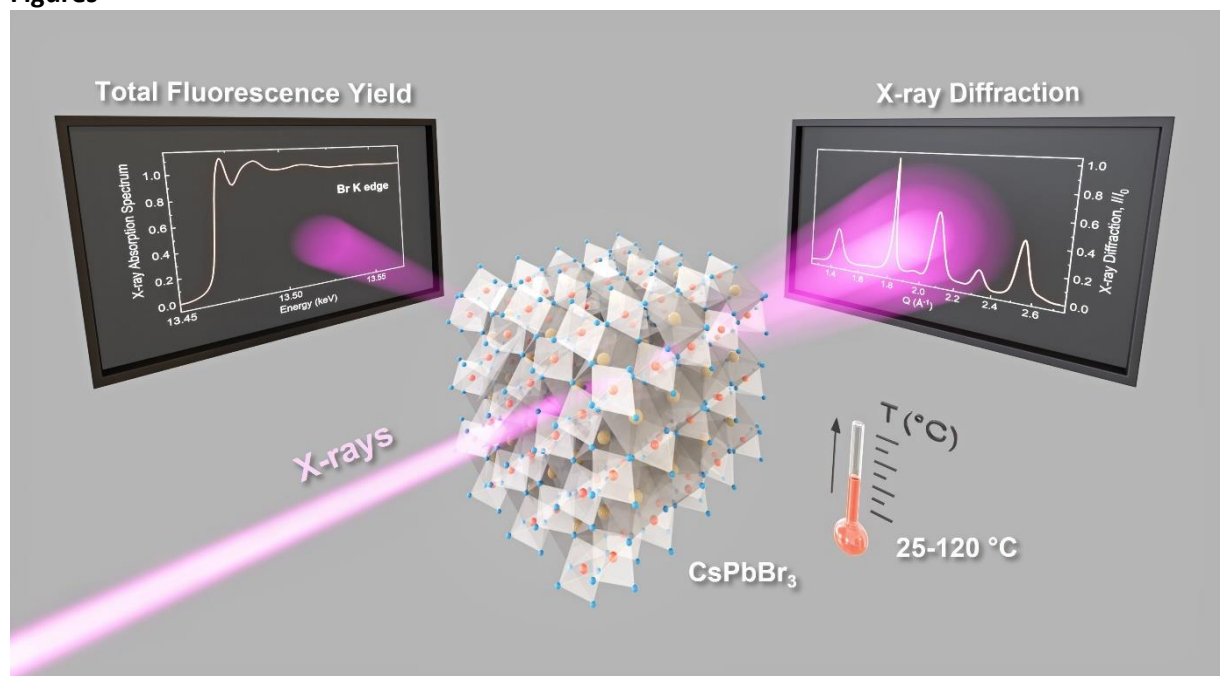

**Figure 1 | Multiscale probing of thermally-induced changes in CsPbBr$_3$ perovskite nanocrystals:** schematic layout of the experiment. Temperature-dependent XRD and XANES measurements were conducted in parallel on CsPbBr$_3$ dry nanocrystals respectively at 12.9 keV and at the Br K-edge (13.450-13.570 keV). Courtesy of Balázs Őrley.

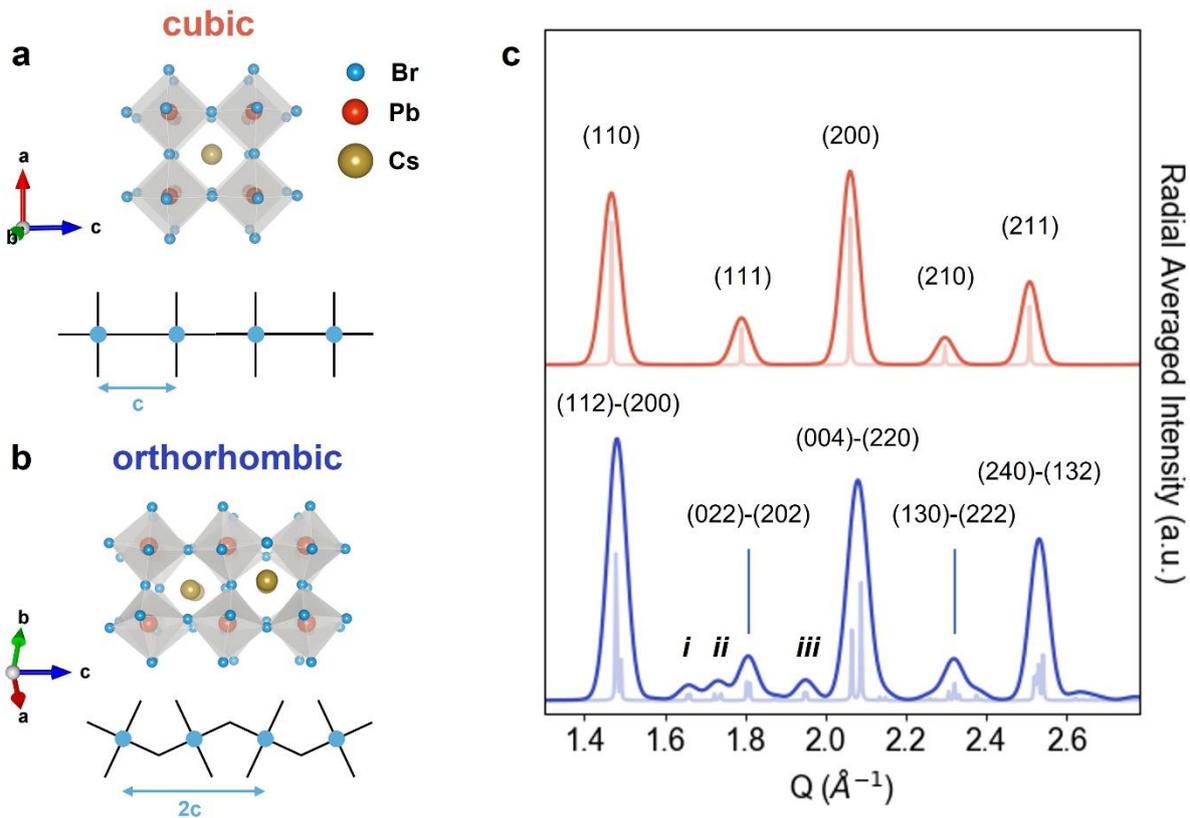

**Figure 2 | CsPbBr₃ crystal structures.** CsPbBr$_3$ **(a)** $Pm\bar{3}m$ cubic and **(b)** $Pnma$ orthorhombic unit cell graphics, each with a schematic for the tilting of the inorganic framework. In the orthorhombic phase, the ordered tilting of the PbBr$_6$ octahedra causes the doubling of the unit cell constant along the crystallographic *c*-axis[44]; **(c)** CsPbBr$_3$ XRD radial averaged intensity plots *I(Q)*. Cubic (top) and orthorhombic (bottom) XRD radial averaged intensity plots *I(Q)*, simulated in VESTA[37] for a 12.9 keV incident X-ray energy. The peaks *(i)*, *(ii)*, *(iii)* correspond to the superlattice peaks and arise from the unit cell doubling of the low symmetry phase with respect to the cubic structure (panel b).

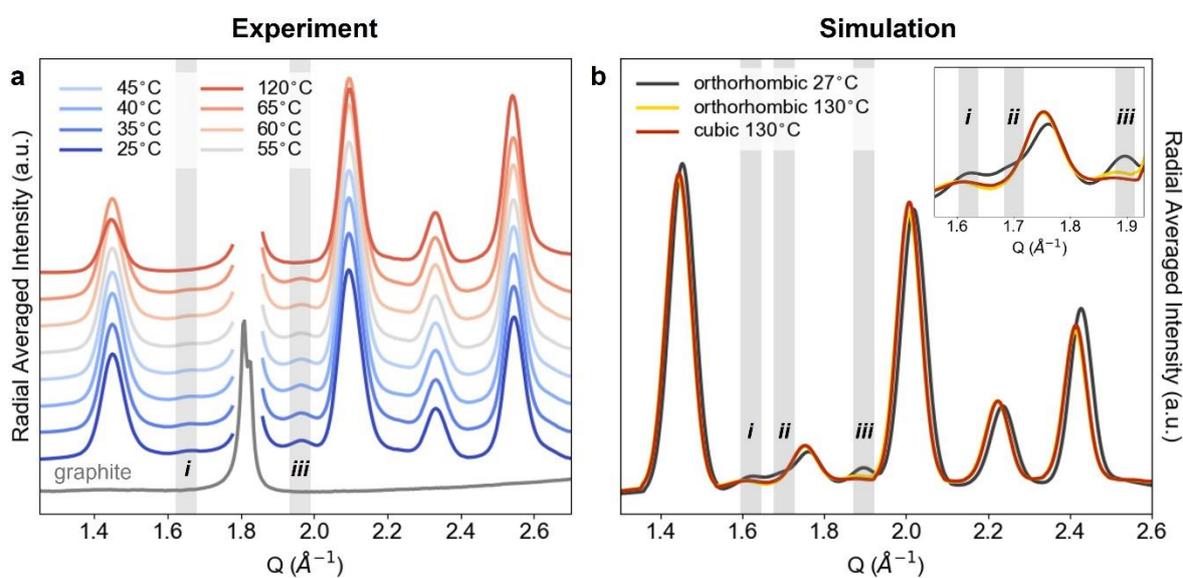

**Figure 3 | Temperature-dependent XRD *I(Q)* profiles. (a)** Experimental XRD *I(Q)* profiles as a function of the temperature from 25 °C to 120 °C. The sharp feature at 1.827 Å$^{-1}$ originates from the graphite peak enclosing the sample. The shaded grey areas mark the region of the *(i)* and *(iii)* superlattice peaks, which disappear upon temperature increase. **(b)** Average XRD *I(Q)* profiles predicted from the MD simulations at: 27 °C with orthorhombic starting geometry (grey), 130 °C with orthorhombic starting geometry (yellow) and 130 °C with cubic starting geometry (red). The shaded grey areas highlight the *(i)*, *(ii)*, *(iii)* superlattice peaks. Inset: zoom into the 1.55-1.94 Å$^{-1}$ region of the superlattice peaks.

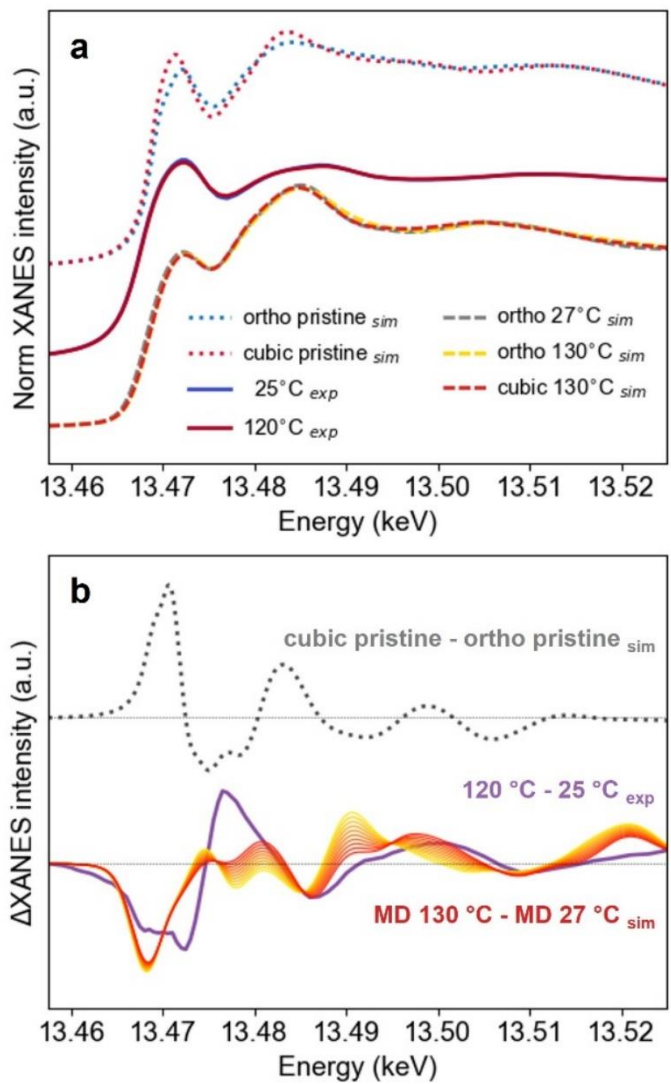

**Figure 4 | Temperature-dependent XANES at the Br K-edge and theoretical ab initio calculations. (a)** Experimental Br K-edge spectra at 25 °C (blue) and at 120 °C (red), computed XANES spectra for the pristine orthorhombic and cubic symmetries (dotted blue and dotted red, respectively) and for the MD simulations at 27 °C (orthorhombic starting symmetry, dashed grey) and 130 °C (orthorhombic and cubic starting symmetries, dashed yellow and dashed red, respectively). All spectra were scaled by their underlying areas and vertically offset. **(b)** Br K-edge XANES differences for 120 °C minus 25 °C (experiment, purple), pristine cubic minus pristine orthorhombic (dotted grey) and the difference between different linear combinations of the cubic 130 °C and orthorhombic 130 °C minus orthorhombic 27 °C MD simulations. The curves were obtained considering different linear combination coefficients of the cubic 130 °C and orthorhombic 130 °C MD curves, from 100% cubic (red) to 100% orthorhombic (yellow). A 3-point adjacent averaging of the energy axis was performed for experimental thermal difference, whereas the simulated spectral differences were multiplied by a factor x0.30 (MD) and x0.15 (pristine), the latter also being vertically shifted, to enable a straightforward comparison with the experiment. In the legend of each panel, "ortho" stands for "orthorhombic".

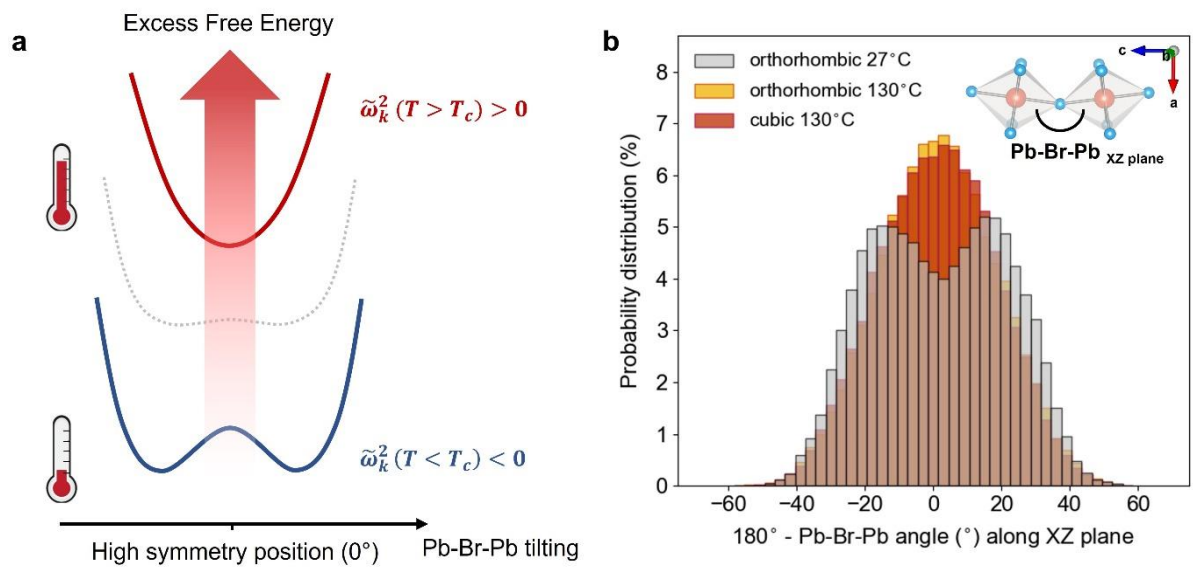

**Figure 5 | Theoretical predictions of the MD simulations. (a)** Schematic of the excess free energy evolution with the temperature along the soft phonon coordinate of the Pb-Br-Pb octahedral tilting. The renormalized frequency of the soft phonon mode $\widetilde{\omega}_k$ is temperature-dependent due to its strong anharmonicity[34]. Upon temperature increase, from bottom to top, the free energy landscape along this mode changes, causing a displacive phase transition. **(b)** Probability distribution (%) of the Pb-Br-Pb angle as a function of the angle distortion: 27 °C (orthorhombic starting geometry, grey), 130 °C (orthorhombic starting geometry, orange) and 130 °C (cubic starting geometry, red). The Pb-Br-Pb angle is projected along the XZ plane, locally describing the octahedral tilting of the Pb-Br inorganic framework. The x-axis reports the difference between 180° and the Pb-Br-Pb angle projection along the XZ plane in order to centre the distribution at 0°, corresponding to the ideal cubic geometry. Upon temperature increase, the Pb-Br-Pb angle probability distribution is modified, changing from bimodal to monomodal across the displacive phase transition. Inset: graphical representation of the Pb-Br-Pb angle in the plane defined by the *a* and *c* crystallographic axes (XZ plane).

**Supporting Information**

# Atomic-level description of thermal fluctuations in inorganic lead halide perovskites

**Table of Contents**

## 1. Samples and characterization

The temperature-dependent X-ray diffraction (XRD) and X-ray absorption near-edge structure (XANES) experiments were performed on a powder of long-chain zwitterion-capped $CsPbBr_3$ dry perovskite nanocrystals (NCs). The sample was prepared according to the literature[1]. The sample specifications are summarized in Table S1. We report the absorption and emission optical spectra (Figure S1a), and the transmission electron microscopy (TEM) image of the NCs (Figure S1b).

| Sample | Absorption Edge | Luminescence Maximum (FWHM) | Quantum Yield (%) | Particle Size (nm) |
|---|---|---|---|---|
| Dry nanocrystals | 503 nm | 511 (22 nm) | 77.4 | 11.9±2.2 |

**Table S1.** Samples characterization for T-dependent XRD and XANES experiments.

## 2. *T*-dependent XRD and XANES data analysis

### 2.1. X-ray diffraction

XRD measurements as a function of the temperature were collected in the temperature range 25-120 °C. The diffraction signal *I(Q, φ)* was recorded using a Pilatus 100k detector (94965 pixels, 172x172 µm² pixel area) placed in transmission geometry at a distance of 24.1 cm from the sample, and displaced vertically with respect to the direct transmitted beam of 4.5 cm. The first order of diffraction was measured in the Q range between 1.3 and 2.8 Å$^{-1}$ using a monochromatized X-ray beam with an energy of 12.9 keV. The signal of the graphite sheets enclosing the powder was isolated in a separate measurement.

The 2D XRD rings were rectified and vertically integrated using the python function pyFAI.AzimuthalIntegrator. In the program, the rotation parameters of the detector were optimized in order to correct the XRD circles into straight lines. The results are shown in Figure S2 for the 25 °C and 120 °C *I(Q)* profiles. A common linear backgorund was subtracted from all XRD curves, obtained from the XRD radial averaged intensity at 35 °C. The pixel-to-scattering vector conversion was determined following the relations:

$$l = p \cdot \Delta$$

$$\theta = \tan^{-1} \frac{l + \alpha}{L}$$

$$d = \frac{\lambda}{\sin \theta}$$

$$S = \frac{2\pi}{d}$$

where p is the pixel number of a given reflection, Δ corresponds to the pixel size, α is the lateral displacement of the detector along the plane orthogonal to the X-ray beam, L is the sample-detector distance and λ corresponds to the X-ray beam energy (12.9 keV) in wavelength. A least-square fitting was employed in order to match the position of the XRD *I(Q)* profile obtained in VESTA[2] for a $CsPbBr_3$ orthorhombic cell, optimizing α and L parameters. The main reflections were attributed following the assignment provided by VESTA for orthorhombic and cubic unit cells.

## 2.2. X-ray absorption near-edge structure

Br K-edge XANES measurements as a function of the temperature were performed in the range 25-120 °C. The spectra were collected in total fluorescence yield geometry, with a 5 element silicon drift detector (SDD) placed at 90° with respect to the incident X-rays. The Br absorption edge was sampled with a non-uniform energy grid of points, denser in the rising edge region (0.5 eV step) and coarse above the edge (2 eV). Systematic energy shifts due to the monochromator backlash across subsequent energy scans were corrected designing an optimization parameter, corresponding to the absolute difference between the modulus of the area underlying the normalized first derivative of the considered spectrum and a reference spectrum (25 °C spectrum). Band gap temperature-dependent contributions to the XANES edge absolute energy position were introduced following Varshni relation[3]. The final spectra were obtained averaging the backlash corrected spectra and the same spectra with the additional Varshni correction, computing the corresponding error bars. For each spectrum, a flat pre-edge offset was subtracted and its intensity was normalized by the edge integral. For a detailed description of the employed procedure, please refer to[4].

## 3. MD computational methods

*Ab initio* MD simulations based on density functional theory (DFT) were performed using the CP2K package[5]. We used the Perdew-Burke-Ernzerhof (PBE) functional[6] to describe the exchange-correlation energy. We employed atom-centred Gaussian-type basis functions to describe the orbitals and an auxiliary plane-wave basis set to re-expand the electron density. DZVP-MOLOPT basis sets[7] were used. The plane wave basis set was defined up to a cutoff energy of 400 Ha. Core-valence interactions were described by Goedecker-Teter-Hutter pseudopotentials[8], with the following valence orbitals: Cs [$5s5p6s$], Pb [$6s6p$], and Br [$4s4p$].

Three different MD simulations, lasting for 10-16 ps and using a timestep of 5 fs, were carried out in the isobaric (*NpT)* ensemble. The external pressure was controlled using the barostat developed by Martyna, Tobias, and Klein[9]. In the runs, the initial shape of the cell was kept constant, while the volume of the cell was allowed to fluctuate. One MD calculation was run at 27 °C (300 K) with the initial orthorhombic geometry. At 130 °C (403 K), two simulations were run, one initialized with the cubic and one with the orthorhombic geometry. The temperature was controlled by a Nosé-Hoover thermostat[10,11]. Simulations were carried out in supercells containing 1080 atoms, which correspond to the 6x6x6 repetition of the unitary cubic cell. The Brillouin zone was sampled at the sole Γ point. The first 5 ps of the simulations were considered as equilibration and were excluded from the statistics. The mean XRD *I(Q)* profiles were calculated averaging the scattering intensities predicted by VESTA[2] for instantaneous structures separated by 0.75 ps extracted from the MD trajectories. Three additional MD simulations, with the same parameters and starting conditions, were also performed for smaller supercells (320 atoms, corresponding to the 4x4x4 repetition of the unitary cubic cell) to generate structures for computationally demanding XANES simulations.

**4. XANES computational methods**

All XANES calculations were performed using the Quantum ESPRESSO distribution[12,13], an open-source code for electronic structure simulations based on DFT and plane-wave and pseudopotentials techniques. The PBE functional[6] was employed to model the exchange correlation effects, and the electron-ions interaction was described using ultrasoft pseudopotentials from the PS-library[14]. Valence states for Br, Pb and Cs atoms were described using semicore 3d, 5d, and 5s plus 5p electrons, respectively. Plane-wave expansions up to a cut-off kinetic energy of 50 and 300 Ry were performed for the Kohn-Sham wave functions and the charge density, respectively. An energy convergence threshold of $10^{-10}$ Ry/atom was defined in the calculations. The Br K-edge absorption cross-section Σ(ω) was computed in the dipole approximation:

$$\Sigma(\omega) = 4\pi^2 \alpha_0 \hbar \sum_{f,\boldsymbol{k},\sigma} \left| \langle \varphi_{f,\boldsymbol{k}}^\sigma | \mathcal{D} | \varphi_{i,\boldsymbol{k}}^\sigma \rangle \right|^2 \delta\left(\varepsilon_{f,\boldsymbol{k}}^\sigma - \varepsilon_{i,\boldsymbol{k}}^\sigma - \hbar\omega\right)$$

where $\alpha_0$ is the fine structure constant, $\varphi_{i,\boldsymbol{k}}^\sigma$ and $\varepsilon_{i,\boldsymbol{k}}^\sigma$ the wavefunction and energy of the initial state, $\varphi_{f,\boldsymbol{k}}^\sigma$ and $\varepsilon_{f,\boldsymbol{k}}^\sigma$ the wavefunction and energy of the final state, and $\hbar\omega$ the energy of the incoming photons. The operator $\mathcal{D} = \boldsymbol{e} \cdot \boldsymbol{r}$ describes the light-matter interaction in the dipole approximation, with $\boldsymbol{e}$ and $\boldsymbol{r}$ being the polarization vector of the incoming light and the electron position, respectively. The Lanczos recursive method was employed to avoid the expensive calculation of the empty states, using the XSpectra package[15,16] of Quantum ESPRESSO. The simulations were performed using the so-called excited-electron plus core-hole (XCH) approach[17]. $\varphi_{i,\boldsymbol{k}}^\sigma$ is the 1s wave function of the emitting Br atom and the finals-state wave functions $\varphi_{f,\boldsymbol{k}}^\sigma$ and $\varepsilon_{f,\boldsymbol{k}}^\sigma$. In this method, one excited Br atom with a full core-hole in the 1s state is included in the supercell and its excited configuration is generated with the LD1 module of the Quantum ESPRESSO package. A charge-neutral excitation is obtained adding an extra electron in the bottom of the conduction band, using a large enough supercell to prevent spurious interactions between the excited atom and its replicas in periodic boundary conditions.

For the pristine orthorhombic and cubic structures, the XANES simulations were performed using 160-atoms supercells with atomic coordinates and lattice parameter extracted from PDF refinements of XRD data recorded at 22 °C and 160 °C, respectively[18]. The supercell Brillouin zone was sampled with a uniform grid of 3x3x2 **k**-points centred in the Γ point. Separate XCH supercell calculations were performed for non-equivalent Br atoms in the structure, with the final XANES spectra resulting from the average of each non-equivalent Br site over three values of the polarization of the incoming light, namely along the [001], [010], [100] crystallographic directions. The spectra were convoluted with an energy-dependent Lorentzian broadening[19], starting with 0.3 eV and reaching the maximum value of 6.0 eV, with an arctan-type behaviour. The inflection point was set 16 eV above the top of the valence band. The convergence of the XANES spectra with the size of the supercell in the presence of a core hole is reported in Figure S3 for the simulations with 160- and 320-atoms supercells. The agreement between the two curves shows that a 160-atoms supercell is sufficiently large to prevent spurious interaction between the core hole replicas in periodic boundary conditions, and was thus used for the XANES calculations of the pristine structures.

More extensive XANES calculations were run using the lattice structure of *ab initio* MD simulations performed with CP2K package[5]. Supercells corresponding to the 4x4x4 repetition of the unitary cubic cell (320 atoms) were considered for three MD calculations, respectively at 27 °C (300 K, orthorhombic starting symmetry) and 130 °C (403 K, both orthorhombic and cubic starting symmetries). The Brillouin zone was sampled only at the Γ point. For each MD simulation, a total of 50 spectra were averaged together, obtained from five different instantaneous structures along the MD trajectory. For each structure, 10 separate calculations were performed placing a core-hole on a randomly chosen Br site, for which the XANES spectrum was computed. In the pseudopotential approximation, the energy scale has no physical meaning since only valence electrons are considered, hence a core-level shift is needed before averaging and comparing spectra from different simulations. The core-level shift for the *i-th* configuration (single MD snapshot and emitter) is taken into account as follows[20]:

$$E \rightarrow E - \varepsilon_{LUB}^i + (E_{XCH}^i - E_{GS}^i)$$

where $\varepsilon_{LUB}^i$ is the energy of the lowest unoccupied band, $E_{XCH}^i$ is the energy of the system with one 1*s* core hole and one electron in the lowest unoccupied state, and $E_{GS}^i$ the energy of the ground state.

Figure S4a shows the XANES spectra of the 50 Br sites considered for the three MD simulations and their averages. The convergence of the spectral shape as a function of the number of Br sites is reported in Figure S4b and was evaluated considering the mean value of the relative (%) standard deviation between each spectrum and the corresponding average:

$$C = < \sqrt{\frac{1}{N} \sum_i^{\# Br\ sites} [I_i(E) - \bar{I}(E)]} >_E \cdot 100 / < \bar{I}(E) >_E$$

where $<>_E$ indicates the average over all the energy points of the XANES spectrum, $I_i(E)$ is the XANES intensity of the *i-th* Br K-edge spectrum at the energy *E*, and $\bar{I}(E)$ is the average XANES intensity at the energy E over all the Br sites. For all MD simulations, the convergence of the XANES spectrum is reached averaging up to 20 Br sites.

The high energy resolution XANES spectra obtained for the simulated MD structures at 27 °C (orthorhombic, black) and 130 °C (orthorhombic and cubic respectively in orange and red) are reported in Figure S5. These spectra include a Lorentzian broadening of 0.5 eV. In Figure 4 (main manuscript), an additional Gaussian broadening was applied to the simulations in order to retrieve an energy resolution comparable to the experiment. In Figure 4 and Figure S5, a rigid shift of the simulated XANES spectra was introduced in order to align the main edge peak to the experimental XANES spectra.

The calculations reported in the manuscript were performed under periodic boundary conditions on sufficiently large supercells in order to simulate a statistically representative ensemble of configurations in a bulk system. No surface atoms were included in the calculations, nor the presence of zwitterionic ligands, being computationally too intensive. The parameters of our simulations were defined to include all physical phenomena necessary to accurately describe the system (thermal effects, configurational- and time-average over multiple local structures, core hole effect on the XANES line shape) with the computational time constrained by the technical feasibility of the calculation.

**5. Time evolution of the Pb-Br-Pb angle distribution in the MD simulations**

In Figure S6 we show the evolution of the Pb-Br-Pb angle distribution projected along the XZ plane as a function of time for the three MD simulations, starting respectively at: (top) 27 °C with an orthorhombic symmetry; (middle) 130 °C with an orthorhombic symmetry; (bottom) 130 °C with a cubic symmetry. Data is reported as the difference between 180° and the Pb-Br-Pb angle projection along the XZ plane in order to centre the distribution at 0°. The simulation at 27 °C shows an initial widening of the angular distribution that completes within the first 200 fs. This effect is due to the thermal dynamics process, which distorts the lattice with respect to the ideal orthorhombic structure imposed in the MD simulation at time zero. In the subsequent MD times, the distribution dynamically fluctuates, retaining the maxima in the symmetric positions around ±16.1°, corresponding to the octahedral tilting cooperatively kept by the PbBr$_6$ units in the orthorhombic structure. For the MD simulations at 130 °C, the initial structure relaxes towards an energetically more stable configuration in longer times. During the first 2 ps, the orthorhombic structure at 130 °C evolves from a bimodal distribution towards a broad monomodal distribution approximately centred at 0°. Instead, the structural relaxation of the cubic structure at 130 °C completes in the first 300 fs of the MD, and it consists in a simple broadening of the distribution due to thermal lattice dynamics. For MD times above 4 ps, the high temperature configurations are analogous, characterized by a Pb-Br-Pb distribution centred in the high symmetry cubic position, but with significant distortions of the octahedral tilting angle.

The time average of the statistical distribution for the three MD simulations is reported in Figure S7 together with the fitting curves. The MD at 27 °C was fitted using two identical Gaussian functions centred at symmetrically distorted angles (±16.1°) and having a standard deviation of 12.0° each. The 130 °C MD simulations initialized with an orthorhombic (cubic) symmetry was fitted using a Gaussian function centred in the high symmetry position of 0° and characterized by a standard deviation of 18.0° (18.2°). The centre of the fitting curves was constrained to 0° and the standard deviation was left as free parameter of the fit. The small deviations between the fitting curves centred at 0° and the histograms (centred at 1.5° and 1.0° for the 130 °C orthorhombic and cubic MD simulations, respectively), confirm that the MD calculations are close to full convergence, which would be achieved with a larger sampling of the MD trajectories.

## 6. Thermal displacements of the Cs, Pb, Br sites in the MD simulations

The thermal dynamics of the Cs, Pb, and Br sites was quantified as a function of the temperature through the probability distribution of the site displacement with respect to their mean position in the MD structure, as reported in Figure S8. The results were averaged over the last 5 ps of the MD simulations, when the structures had relaxed into the configurations of minimum energy. We observe that the Cs fluctuations are significantly active already at room temperature (<$\Delta r_{Cs}^{27°C}$>=0.62 Å) and they increase with temperature (<$\Delta r_{Cs}^{130°C}$>=0.73 Å, +18% upon temperature increase). This result confirms the wide mobility of the inorganic A$^+$ cations in their cuboctahedral voids[21]. The Pb thermal dynamics is only slightly perturbed by the temperature increase ($\Delta r_{Pb}^{27°C}$=0.33 Å, $\Delta r_{Pb}^{130°C}$=0.38 Å). Instead, the Br centres exhibit significant changes in their mobility with temperature, showing that the thermal lattice dynamics of the Pb-Br framework is dominated by the Br displacements ($\Delta r_{Br}^{27°C}$=0.59 Å, $\Delta r_{Br}^{130°C}$=0.73 Å, i.e. +24% upon temperature increase). These observations highlight the increased lattice mobility associated with the phase transition from room temperature to 130 °C.

**SI Figures**

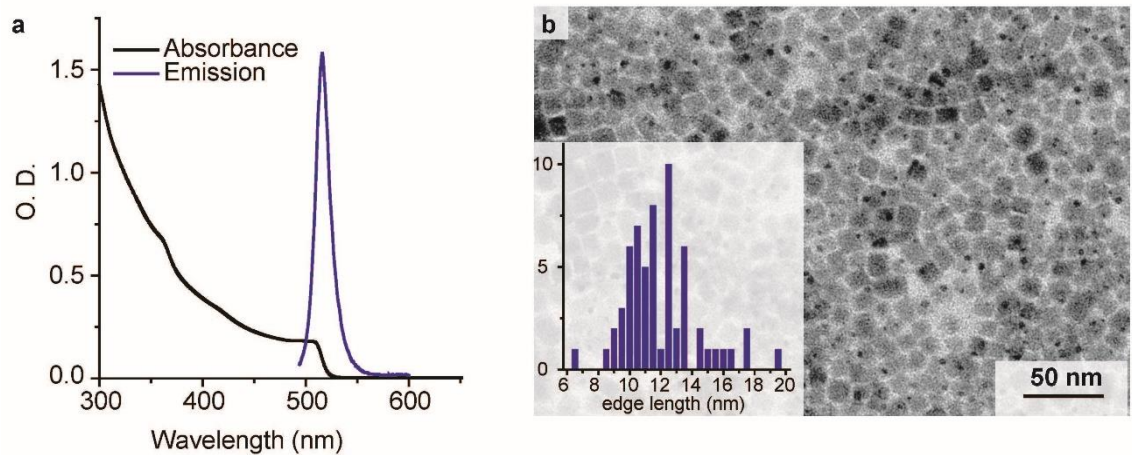

**Figure S1. (a)** Absorption and Emission Optical Spectra and **(b)** TEM image of the CsPbBr$_3$ NCs. The inset in **(b)** shows the distribution of the cuboidal NCs average edge length (nm), with the y-axis corresponding to the number of counts.

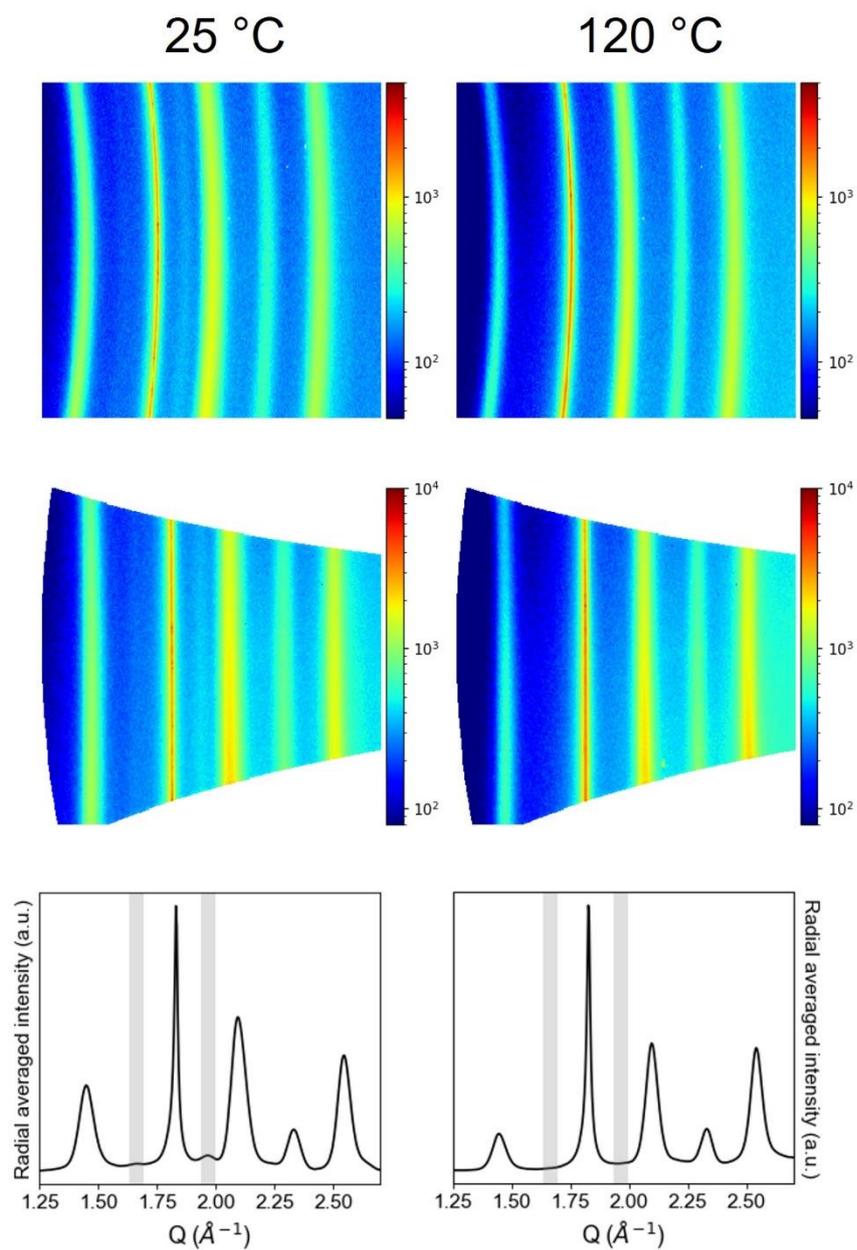

**Figure S2 | XRD results for 25 °C (left) and 120 °C (right)**: powder diffraction rings measured on the 2D detector (top), rectified 2D diffraction patterns upon data analysis (middle), and 1D radial averaged intensity plots *I(Q)*, obtained by vertically binning the rectified maps (bottom).

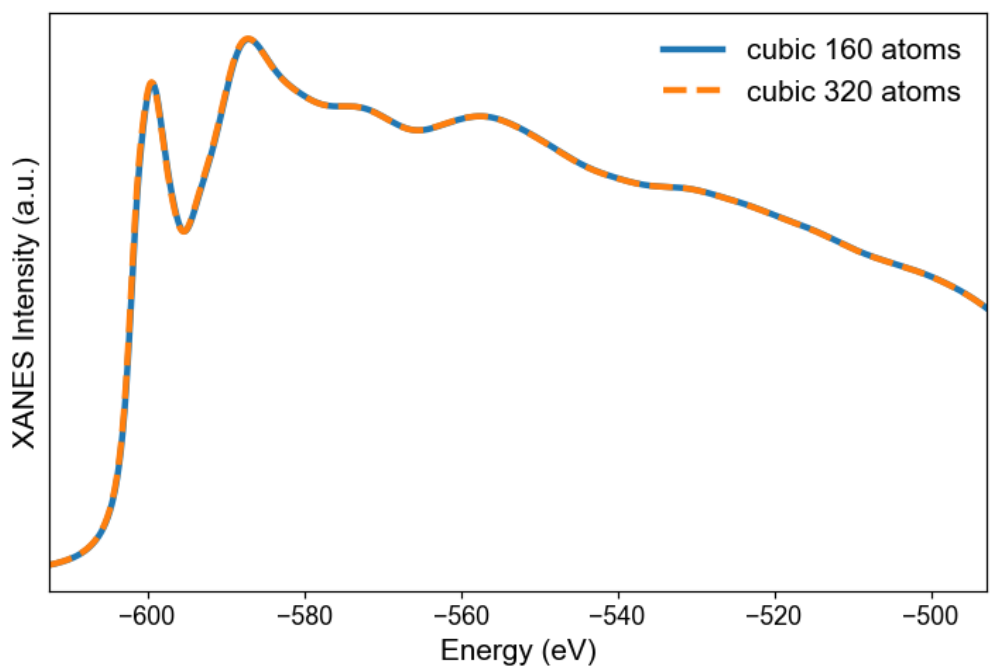

**Figure S3 | Computed XANES spectra for pristine cubic structures in the presence of a core hole:** XANES simulations performed using 160-atoms (blue) and 320-atoms (dashed orange) supercells. The Br K-edge spectra were calculated for the Br site on which the core hole was localized.

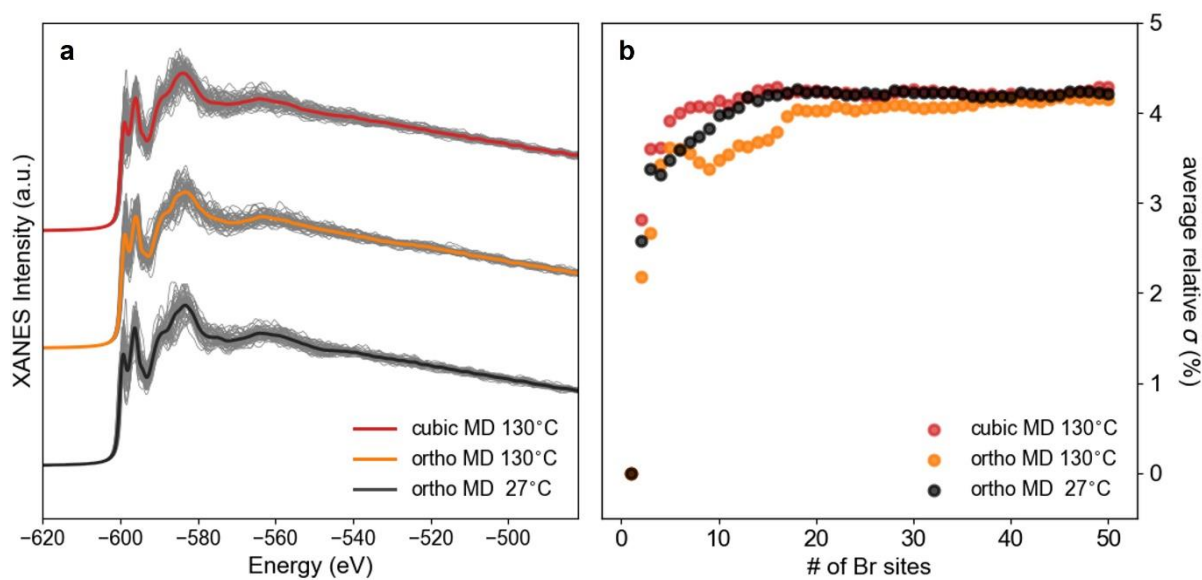

**Figure S4 | Computed XANES spectra from MD simulated structures: (a)** XANES spectra for single Br sites (thin lines) and their average (thick lines) for the MD structures at 27 °C (orthorhombic, black) and 130 °C (orthorhombic and cubic, in orange and red, respectively); **(b)** convergence parameter (average relative % standard deviation of the XANES intensity) as a function of the number of Br sites for the XANES simulations of the orthorhombic 27 °C (black), orthorhombic 130 °C (orange) and cubic 130 °C (red) MD trajectories. In the legend of each panel, "ortho" stands for "orthorhombic".

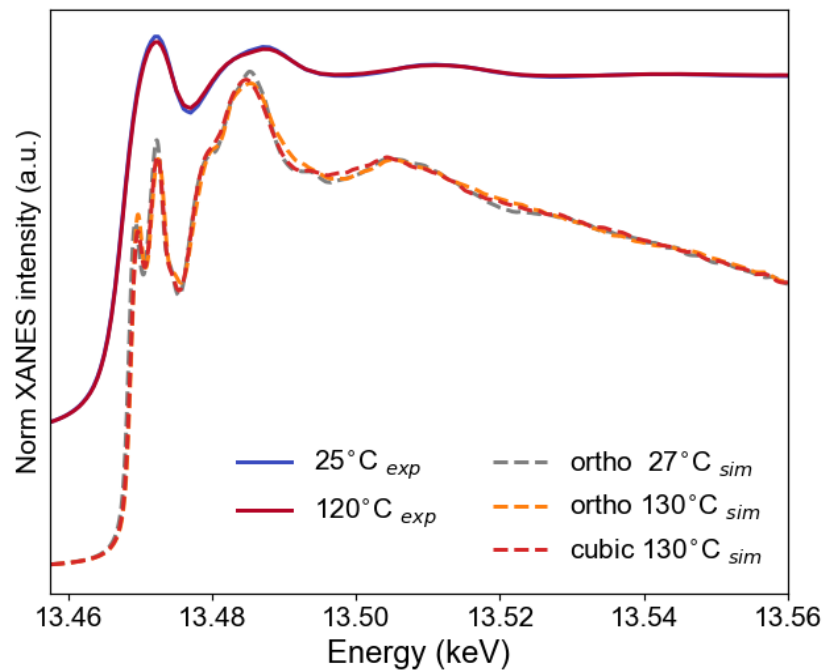

**Figure S5 | Comparison of experimental and high energy resolution MD XANES Br K-edge spectra:** experimental XANES spectra at 25 °C (blue) and 120 °C (red) and computed high energy resolution spectra for the MD structures at 27 °C (orthorhombic, black) and 130 °C (orthorhombic and cubic in orange and red, respectively). All spectra were scaled by their underlying areas and a vertical offset was applied to the experimental traces. In the legend, "ortho" stands for "orthorhombic".

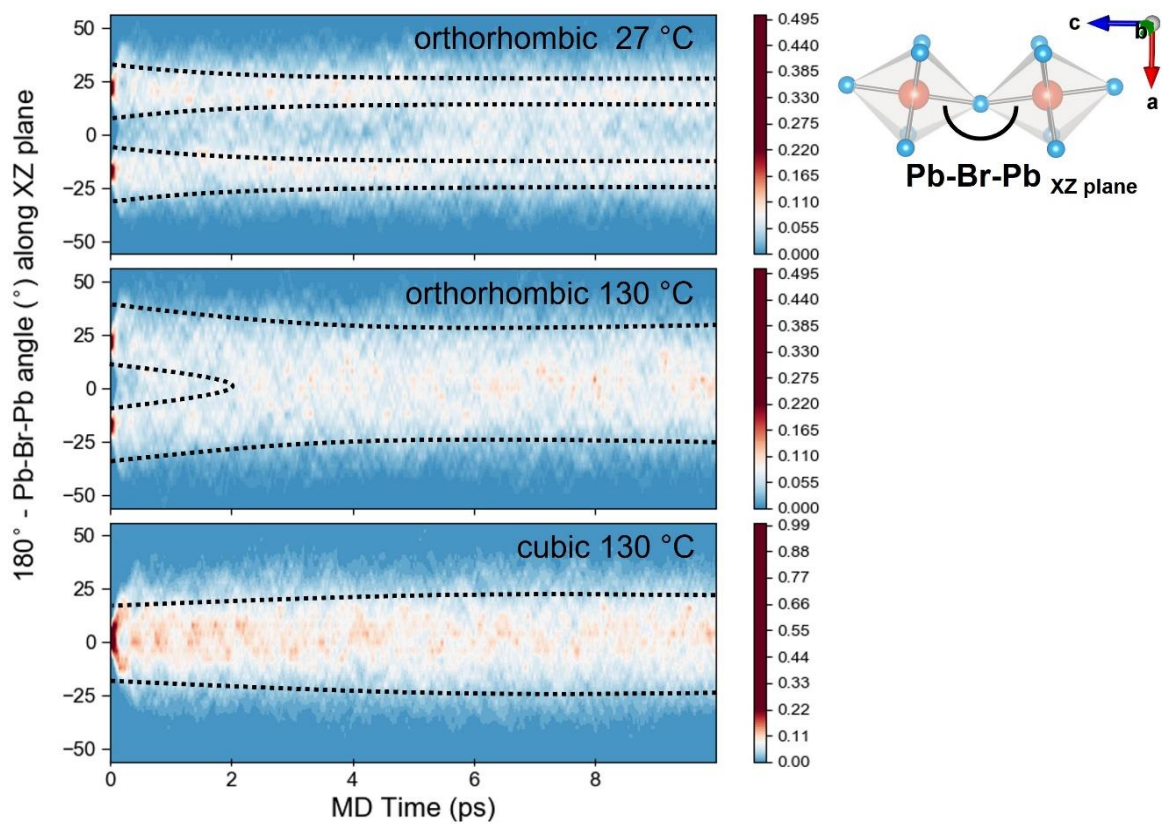

**Figure S6 | Time evolution of the Pb-Br-Pb distribution along the XZ plane during the MD simulations**. MD at the following starting conditions: (top) 27 °C from an orthorhombic crystal structure; (middle) 130 °C from an orthorhombic crystal structure; (bottom) 130 °C from a cubic crystal structure. The y-axis reports 180° minus the Pb-Br-Pb angle projection along the XZ plane in order to centre the distribution at 0°. The color-code shows the probability to find a given angle for the Pb-Br-Pb projection on the XZ plane as a function of the time. The dark dashed lines serve as guides for the eyes.

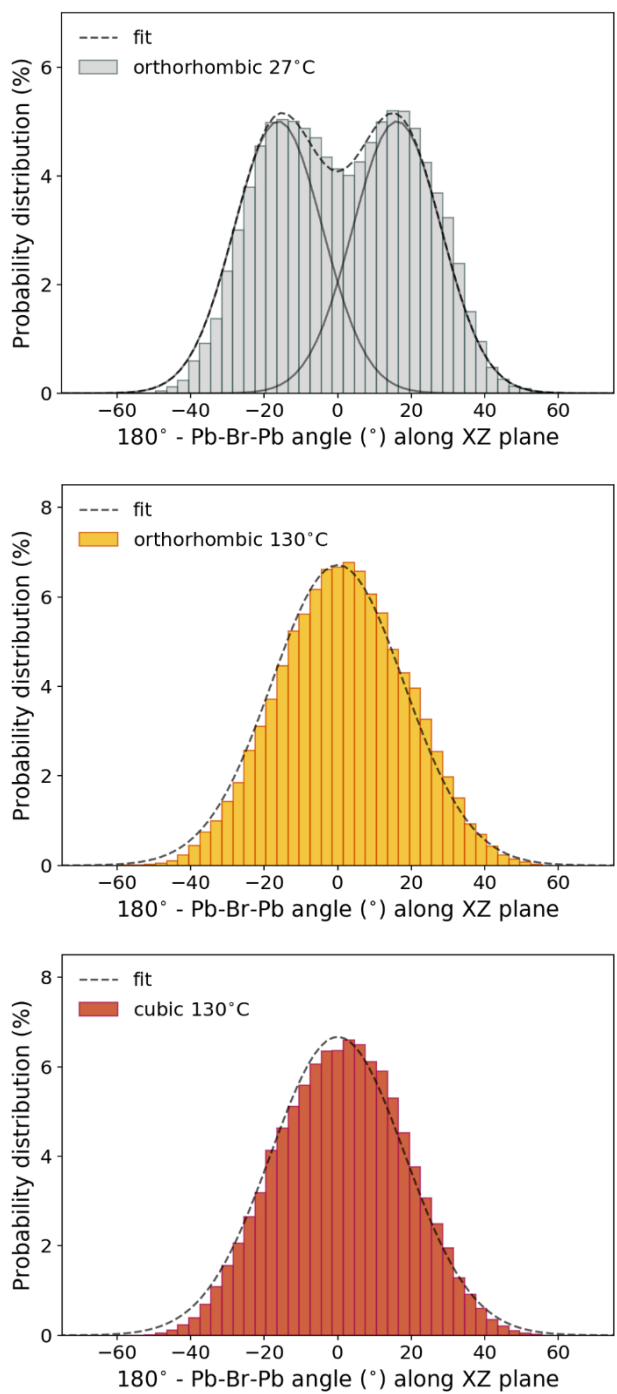

**Figure S7 | Probability distribution (%) of the Pb-Br-Pb angle as a function of the angle distortion.** MD with the following starting conditions: (top) 27 °C with orthorhombic geometry; (middle) 130 °C with orthorhombic geometry; (bottom) 130 °C with cubic geometry. The corresponding Gaussian fits are reported in grey (dashed line). The y-axis reports 180° minus the Pb-Br-Pb angle projection along the XZ plane in order to centre the distribution at 0°. The dashed curves correspond to the symmetrized Gaussian fit of the probability distributions.

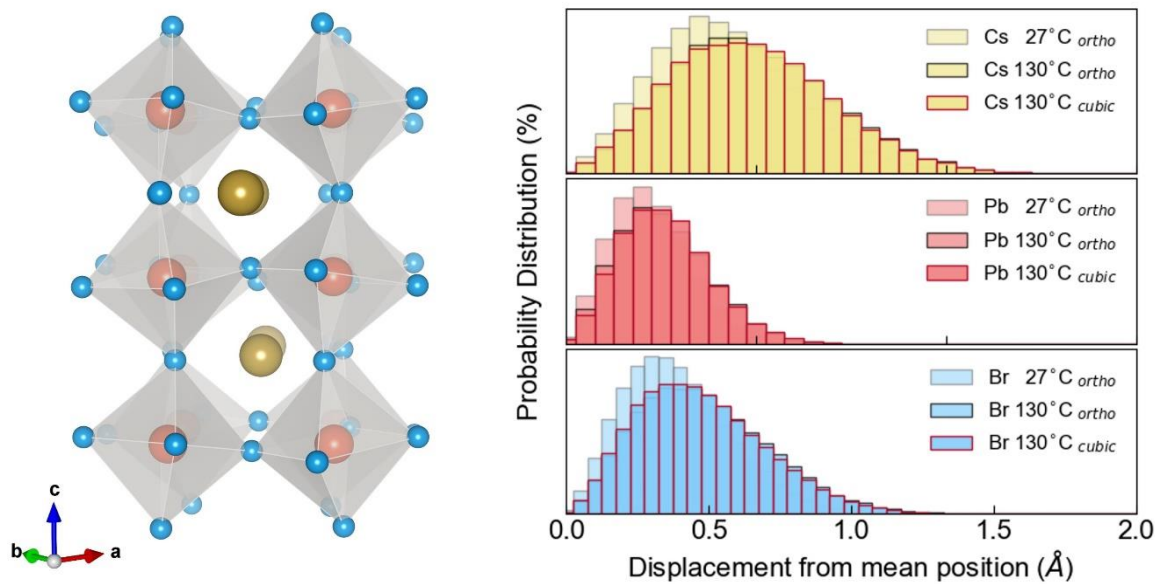

**Figure S8 | Theoretical predictions of the MD simulations.** Probability distribution (%) of site displacement (Å) from the average position over the last 5 ps of the MD simulations, in the three starting cases: 27 °C orthorhombic geometry (grey boxing), 130 °C orthorhombic geometry (black boxing), 130 °C cubic geometry (red boxing). Top - Cs atom (yellow); middle - Pb atom (light red); bottom - Br atom (light blue). The atoms are color-coded with the graphical representation of the orthorhombic structure reported on the left. In the legend of each panel, "ortho" stands for "orthorhombic".